\documentclass[11pt]{article}
\usepackage[cp1251]{inputenc}
\usepackage[english]{babel}
\usepackage{graphicx}
\usepackage{amssymb}
\usepackage{amsmath}
\usepackage{amsthm}
\usepackage{amsfonts}
\usepackage{amssymb}
\title{{\bf \Large  Fractional-order derivatives in cosmological models of accelerated expansion}\\
{\normalsize ~~{\bf V.\,K. Shchigolev}\thanks{E-mail:
vkshch@yahoo.com}}\\
{\small {\it Ulyanovsk State University, 42 L. Tolstoy Str.,
Ulyanovsk, 432000, Russia}}\\
\vspace{2mm}
\small \begin{quote}{\bf Abstract} --  In this brief review, we present the results of the fractional differential approach in cosmology in the context of the exact models of cosmological accelerated expansion obtained by several authors to date. Most of these studies are devoted to the problem of introducing fractional derivatives or fractional integrals into the classical General Relativity (GR). There are several observational and theoretical motivations to investigate the modified or alternative theories of
GR. Among other things, we cover  General Relativity modified by a phenomenological approach dealing with fractional calculus. At the same time, a sufficiently large number of exact solutions of the cosmological equations modified by this approach were obtained. Some of these models may be especially relevant in the light of solving the problem of late accelerated expansion of the universe. These studies are largely motivated by rapid progress in the field of observational cosmology that now allows, for the first time, precision tests of fundamental physics on the scale of the observable Universe. The purpose of this review is to provide a reference tool for researchers and students in cosmology and gravitational physics, as well as a self-contained, comprehensive and up-to-date introduction to the subject as a whole.
 \\
\vspace{2,5mm}
{\bf PACS numbers}: 98.80.-k; 04.50.Kd; 98.80.Jk; 04.20.Jb. \\
{\bf Key words}: Fractional Derivative and Integral, Cosmological models, Exact solutions, Accelerated expansion.\\
\end{quote}}
\date{}

\begin{document}

\maketitle
\vspace{-2.5cm}
\section{Introduction}

The well-known cosmological observations, including a Type Ia supernova, cosmic microwave background radiation, and large-scale structure, strongly indicate that our universe is undergoing a phase of late accelerated expansion \cite{Riess1}-\cite{Allen}.

The fast rising of the number of publications over the past two decades, devoted to the mysterious late cosmological acceleration, led to a radical revision of cosmological models. However, two main directions of such modifications should be distinguished. First of all, it has been developed a lot of models with different exotic forms of matter, often called Dark Energy (DE), which satisfies the equation of state (EoS) $ w = p / \ rho <-1 / 3 $. Different types of the DE models have been proposed, such as the cosmological constant\cite{Peebles}, quintessence \cite{Caldwell1,Sami},  phantom \cite{Caldwell2}-\cite{Cline}, tachyon \cite{Sen1, Gibbons}, k-essence \cite{Mukhanov}-\cite{Scherrer}, Chaplygin gas \cite{Kamenshchik}, quintom \cite{Feng}, holographic dark energy \cite{Horava} and Yang--Mills fields \cite{0Shchigolev, 00Shchigolev}, etc.

An alternative approach to the problem of the late accelerated expansion of the Universe is associated with the consideration of numerous modifications of the theory of gravity itself. Various modifications include multidimensional theory, brane world models, teleparallel theory and many others \cite{Sahni1} - \cite{Brax}.  Moreover, we could recall the broad class of modified gravity, such as $ f (R) $, $ f (G) $, $ f (R, G) $, $ F (R, T) $, etc., which produce adequate gravitational alternatives for DE \cite{Starobinsky1} - \cite{Myrzakulov}.
Nevertheless, the mystery of the late cosmological acceleration is still one of the main problems of modern cosmology.

At the same time, taking into account the cosmological inflation paradigm of the very early Universe, any reliable cosmological model should consist of at least two periods of accelerated expansion. Therefore, it is interesting to consider any cosmological models that are able to more or less realistically describe the entire evolution from the beginning to the present day \cite{Nojiri1} -\cite{4Shchigolev}.

Beyond  to such modifications, a certain phenomenological approach in theoretical cosmology can be noted in connection with the problem of late cosmological acceleration.
For example, in Refs. \cite{Dvali1}-\cite{Dvali2},  the massive gravity theory is considered as a modification of GR and include massive gravitons. This theory has a rich phenomenology, such as explaining the accelerated expansion of the universe without invoking dark energy, and have attracted much attention recently \cite{Clifton}.

A class of phenomenological models based on the ideas of fractional calculus seems to be interesting and attractive in its attempts to describe the periods of accelerated expansion. The cosmological equations describing the dynamics of a homogeneous and isotropic Universe are the systems of ordinary differential equations.
Modifying the set of cosmological equations in this appraoch0, one should avoid the conflict between the integer dimension of (pseudo) Riemann space - time of GR and
fractional order of derivatives in the modified equations.  As it emphasized in Refs. \cite {C5, 1Shchigolev}, there are two different methods of such modification towards the  fractional derivative cosmology. The
simplest method is the Last Step Modification (LSM) method, in which the
Einstein's GR equations  are replaced
with  the fractional analogous. In other words, the substitution $
\partial _t\to D_t^\alpha$, where  $D_t^\alpha$ is a derivative of fractional order $\alpha$,  should be done after the field equations for a
specific geometry have been derived. The fundamentalist
methodology could be the First Step Modification (FSM), in which one
starts by constructing fractional derivative geometry. The intensively
developed approach to modification of the main cosmological
equations and non-conservative systems of Lagrangian dynamics on
the basis of a variational principle for the action of a
fractional order (Fractional Action-Like Variational Approach
FALVA) developed in \cite{C2}-\cite{C4} represents one of the
possible version of intermediate modification (Intermediate Step
Approach, ISA) mentioned in \cite{C5}.

The main part of articles, which have been published by now, devotes to the fractional-order derivatives cosmology on the bases of FSM, ISA, LSM and FALVA. In ISA and FALVA,  the set of  cosmological equations can be readily  obtained by the direct replacing derivatives or using a variational principle for the Einstein-Hilbert action.  In any case it is easy to generalize
masters equations of the field theories to the case of fractional derivatives. Therefore, one could easily  obtain the fractional-differential analogue of the Friedmann equations even in the case of involving  some physical field as a source of gravitation in frameworks of ISA  as well as on the basis of FALVA.

The paper is organized as follows. In Section 2, we briefly review the cosmological models with fractional-order derivatives. In Section 3, we address the fractional Einstein-Hilbert action cosmological models.  In Section 4, we make some conclusive remarks. Finally, in Appendix A, we present a brief information concerning the definitions and some features of  fractional  derivatives and integrals.

\section{Studying cosmological models with fractional-order derivatives}

Since the appearance of the first cosmological models with fractional derivatives, this direction of research has not received proper development. Consider what we have here by now.

\subsection{Cosmological models with fractional derivatives}

Presumably the first cosmological models with equations containing fractional derivatives were proposed and discussed in \cite {C2}. In order to avoid a conflict between the
occurrence of fractional derivatives in the Friedmann equations and
classical definition of a tensor in the Einstein equation, based on
the integer order derivative in tensor law of transformation, the
quoted author has repeated the well known derivation of the Friedmann
equations for a dust from the classical approach (see, for instance,   \cite {C3},
\cite {C12}) and than has replaced all integer derivatives with
its fractional analog. For the spatially flat Universe, the Friedmann
equations with a cosmological term  are written down in \cite {C2}
as follows:
\begin{equation}
\label{1}
\begin{array}{rcl}
(D^{\alpha}_t a(t))^2=(A_1 (G \rho)^{\alpha}+B(\Lambda
c^2)^{\alpha})a^2,~~\\
{}\\
D^{\alpha}_t(D^{\alpha}_t a(t))=-(A_2 (G \rho)^{\alpha}-B(\Lambda
c^2)^{\alpha})a,
\end{array}
\end{equation}
where $a(t)$  is a scale factor of the Friedmann-Robertson-Walker
(FRW) line element,
\begin{equation}
\label{2} d s^2 = d t^2- a^2 (t)(d r^2+\xi^2 (r)d \Omega ^2),
\end{equation}
where $\xi (r)=\sin r,r,\sinh r$ for the sign of space curvature
$k=+1,0,-1$, consequently. The occurrence of $ \alpha $ - degrees
in the right-hand side of equations (\ref{1}) is caused,
probably, by dimensional reasons. Then the author of paper \cite {C2}
obtained the following  solution to this set in a static case
$a=a_0= constant$:
$$
(G\rho)^{\alpha}=\frac{C_1}{t^{2\alpha}},~~(\Lambda
c^2)^{\alpha}=\frac{C_2}{t^{2\alpha}}.
$$
Then, the conclusion is followed that the density of matter and
cosmological constant decrease as $1/t^2 $ if $G $ and speed of
light in vacuum $c $ remains constant, but if the density of
matter and $\Lambda$ remain constant, then the following holds:
$G\sim 1/t^2 $ and $c\sim 1/t $. The solution of equations (\ref{1}), mentioned as an illustration of the method of \cite{C2}, in
the form $a (t) =a_0 E _ {1, \alpha} (Ct ^ {\alpha}) $, where

$$
E_{\alpha,\beta}(t)=\sum^{\infty}_{k=0}\frac{t^k}{\Gamma(\alpha
k+\beta)}
$$
is the so-called Mittag-Leffler two-parametric function \cite{C1}, is
not confirmed by any calculations.

It is interesting to note that for $k=0 $ and in the absence of matter
and the cosmological constant, that is if $ \rho=0 $ and $ \Lambda =
0 $, equations (\ref{1}) are reduced to
\begin{equation}
\label{3}
D^{\alpha}_t a(t)=0,~~D^{\alpha}_t(D^{\alpha}_t
a(t))=0,
\end{equation}
which for $ \alpha = 1 $ gives the obvious result: $a (t) =a_0
=constant$, and interval (\ref{2}) is reduced to those of the
Minkowski space. If $ \alpha \ne 1 $, then the substitution of
the zero constant (that is the derivative $D ^ {\alpha} _t a (t) $)
from the first equation of (\ref{3}) into the second one results
in identity irrespectively to the definition of fractional
derivative, RLD or CD. The solution of the first equation in (\ref{3}) for CD, as well as in the case of integer derivative, is
equal to constant, but for RLD its partial solution is zero, and
the general solution depends on time: $a (t) \sim t ^ {\alpha -1}
$ for $ 0 < \alpha \le 1 $ (see \cite {C1}, p. 216). This
circumstance and also the fact, that initial conditions for
equations with CD should be expressed by means of integer
derivative, instead of fractional one, as in the case of RLD,
frequently decline a choice of definition for the benefit of CD.
In general considerations, we shall not specify a type of
fractional derivative so long as it will be possible.

It is useful to note that  in \cite {C5} the fractional-derivatives analogue of Friedmann    equations
are written down in other form, namely:
\begin{equation}
\label{4} 3 [ k +(D^{\alpha}_t a)^2]= \kappa\rho a^2, ~~a^3
D^{\alpha}_t p = D^{\alpha}_t [a^3 (\rho + p)],
\end{equation}
where we use $ \rho $ for the energy density and $p$ for the
pressure. The second equation represents the energy-momentum
conservation law ($T^i _ {j; i} =0 $) for the perfect fluid in
space-time (\ref{2}) with the integer-order derivatives replaced
by its fractional analogues. In  the assumption of $k=0$ and the power-law dependence of $a, p $ and $ \rho $ on the time: $a=C  t^n, ~p=A t^m, ~ \rho =B t^r $, the relations between $m, r, n, \alpha, B$ and barometric coefficient $w$ in  the equation of state $p = w \rho $ have been found in paper \cite{1Shchigolev}.

However, the paper \cite{C6} has already been done using the naive approach method \cite {C5}, in which
the Riemann curvature tensor and the Einstein
tensor are defined by the inconceivable Christoffel symbols containing
fractional (of order $ 0 < \alpha \le 1 $) derivatives of metrics
coefficients,
\begin{equation}
\label{5} \Gamma^{\mu}_{\nu \lambda}(\alpha)=\frac {1}{2}g^{\mu
\nu}(\partial^{\alpha}_{\nu}g_{\rho
\lambda}+\partial^{\alpha}_{\lambda}g_{\rho
\nu}+\partial^{\alpha}_{\rho}g_{\nu \lambda}),
\end{equation}
where $ \partial^{\alpha}_{\nu}$ is a fractional derivative
(A.2)  with respect to $x^{\nu}$. So it allowed to write down the fractional analogous for the Einstein equation,
$$
R_{\mu \nu}(\alpha)- \frac{1}{2} g_{\mu \nu}
R(\alpha)= \frac{8 \pi G}{c^4} T_{\mu \nu}(\alpha),
$$
and the geodesic equation,
$$
 \frac{d^2 x^{\mu}}{d \tau^2}+ \Gamma^{\mu}_{\nu
\lambda}(\alpha)\frac {d x^{\nu}}{d \tau}\frac {d x^{\lambda}}{d
\tau}=0.
$$
Unfortunately, the author of \cite{C6} only shows that
linearized equations in the limit $ \alpha \to 1 $ are reduced to
the usual equations with the integer derivatives for the
gravitational waves and the Newtonian potential. That could be predicted
from the very beginning due to the definition of a fractional
derivative. Let us note, that an attempt to prove such fractional
derivative replacement in Christoffel symbols, such as (\ref{5}), is undertaken in \cite{C5} on the basis of fractional-differential geometry, that would be related to FSM
formalism. However, the  author of \cite{C6} also mentioned that it is so far
not clear what such geometry should be. Nevertheless,
an attempt of construction of fractional geometry with the help
of fractional coordinate transformations $d x^i = D^{\alpha}_jx^i
d x^j$ is undertaken  for the flat two-dimensional space in \cite{C5}.

Much more advanced and proved results concerning FSM formalism were obtained by S. Vacaru (see \cite {C13}-\cite {C7} and the
bibliography therein). In those works, the results of construction
of the fractional theory of gravitation for the space - time of
fractional (not integer) dimension are obtained. The author sees
one of the simplest motivation for application of fractional
differential calculus in the theory of gravitation in an
opportunity to avoid singularities of the curvature tensor of
physical meaning due to the completely different geometrical and
physical solutions of the fundamental equations. Besides, it is
noted that models of fractional order are more adapted to
the description of processes with memory, branching and hereditarity,
than those of integer order. The result of the application of the
method developed by the author of nonholonomic deformations to
cosmology was the construction of new classes of cosmological
models \cite {C14}.

\subsection {Fractional derivatives cosmology with a scalar field}

First, the naive (or LSM) approach to the
fractional derivative cosmological models of a scalar field has been considered in Ref. \cite{1Shchigolev}. Following LSM, the  substitution of
fractional derivative of the scale factor and the scalar field
instead of integer-order derivatives in GR equations yields:
\begin{eqnarray}
D^{\alpha}_t(D^{\alpha}_t \phi) +3\left(\frac{D^{\alpha}_t a}{a}\right)
D^{\alpha}_t \phi + \frac{d V(\phi)}{d\phi}=0{~,}~~~~~~\label{6}\\
\left(D^{\alpha}_t a\right)^2+k=\frac{ 8\pi
G}{3}\left(\frac{1}{2}\left(D^{\alpha}_t
\phi\right)^2+V(\phi)\right)a^2+
\frac{\Lambda}{3}a^2{~,}\label{7}\\
D^{\alpha}_t(D^{\alpha}_t a)= -\frac{8\pi
G}{3}\left(\left(D^{\alpha}_t
\phi\right)^2-V(\phi)\right)a+\frac{\Lambda}{3}a{~.}~~~\label{8}
\end{eqnarray}
As known, among three equations of standard GR
cosmology (when $\alpha = 1$) only two equations are independent. In the case  $\alpha \neq 1$, all three equations of (\ref {6}) - (\ref {8}) are generally
independent  due
to the modified Leibniz rule for the fractional derivative \cite{C1}.
Because the solution of the nonlinear fractional equations (\ref {6}) - (\ref{8}) is much complicated compared to their classical prototype, an example of an exact solution to the equations (\ref {6}) - (\ref {7})  is given in Ref. \cite{1Shchigolev} to demonstrate the existence of exact solutions for such a model.

Considering further the models followed from the intermediate approach (or ISA) in \cite{1Shchigolev}, the main equations have been derived  from the variational
principle for the Einstein-Hilbert action according to  ADM
formalism in cosmology (see, e.g. , \cite{C22}). For this purpose,  the derivatives over time in the Einstein-Hilbert action $S_{EH}\equiv\int L_{EH} dt$ for FRW model of the
Universe, filled with a real homogeneous
scalar field $\phi (t)$,
\begin{equation}
\label{9} d s^2 = N(t)^2 d t^2- a^2 (t)(d r^2+\xi^2 (r)d \Omega
^2),
\end{equation}
where $N$ is a laps function, have been replaced with
its fractional (of order $ \alpha $) analogous. The result is as follows:
\begin{equation}
\label{10} S_{EH}=\int dt N
\left[\frac{3}{8\pi G} \left( -\frac{a (D^{\alpha}_t
a)^2}{N^2}+ka-\frac{\Lambda a^3}{3}\right) +  a^3
\left(\frac{(D^{\alpha}_t \phi)^2}{2N^2}-V(\phi)\right)\right].
\end{equation}
where $V(\phi)$ is a potential of the field.
Variational problem with fractional derivatives for functions $q_j
(t) $ in the action $S [q_j] (t) = \displaystyle\int\limits _ {c}
^ {d} L (q_j (t), {} _cD _ {t} ^ {\alpha} q_j (t), {} _t D _ {d} ^
{\beta} q_j (t)) d t $ on the interval $[c, d]$ yields  the
modified Euler-Lagrange equations \cite{C1},\cite{C24}:
\begin{equation}
\label{11} \frac{\partial L}{\partial q_j} + {}_t D_{d}^{\alpha}
\left(\frac{\partial L}{\partial ({}_c D_{t}^{\alpha}q_j)}\right)
+ {}_c D_{t}^{\beta} \left(\frac{\partial L}{\partial ({}_t
D_{d}^{\beta}q_j)}\right) = 0.
\end{equation}
In the case of (\ref{10}), one has $q_j (t) = \phi (t), N (t) $ and $a (t) $. Therefore
for $L _ {EH} $ from (\ref{10}) and (\ref{11}),  the
following basic equations of the model can be derived:
\begin{eqnarray}
{}_t D^{\alpha}(a^3 D^{\alpha}_t \phi) -
a^3 \frac{d V(\phi)}{d\phi}=0{~,}~~~~~~~~~~~~~~~~~~~~~~~\label{12}\\
\left(D^{\alpha}_t a\right)^2+k=\frac{ 8\pi
G}{3}\left(\frac{1}{2}\left(D^{\alpha}_t
\phi\right)^2+V(\phi)\right)a^2+
\frac{\Lambda}{3}a^2{~,}~~~~~~~~~\label{13}\\
2 {}_t D^{\alpha}(a~D^{\alpha}_t a)+\left(D^{\alpha}_t
a\right)^2-k=8 \pi G \left(\left(D^{\alpha}_t
\phi\right)^2-V(\phi)\right)a^2-\Lambda a^2{~,}\label{14}
\end{eqnarray}
where  ${}_0 D^{\alpha}_t \equiv D^{\alpha}_t$ and ${}_t
D^{\alpha}_{\infty} \equiv {}_tD^{\alpha}$.

In the case of the limit ($ \alpha \to 1 $),  one has $D^1_t
\to \displaystyle\frac {d} {dt} $, $ {} _t D^1 \to -\displaystyle\frac
{d} {dt} $, and the set of fractional cosmological equations  (\ref{12}) -(\ref{14}) tends to the classical set of the Friedmann and scalar field equations:
\begin{eqnarray}
\ddot \phi+3\frac{\dot a}{a}\dot \phi+\frac{d V(\phi)}{d\phi}=0,~~~~~~~~~~~~~~~\label{15}\\
2\frac{\ddot a}{a}+\frac{\dot a^2}{a^2}+\frac{k}{a^2} = -8\pi
G\left(\frac{\dot\phi^2}{2}-V(\phi)\right)
+\Lambda,\label{16}\\
\frac{\dot a^2}{a^2}+\frac{k}{a^2}=\frac{ 8\pi
G}{3}\left(\frac{\dot\phi^2}{2}+V(\phi)\right)+\frac{\Lambda}{3},~~~~\label{17}
\end{eqnarray}
which had to be expected from the very beginning.

\subsection{The recent studies in fractional derivatives gravity and cosmology}

As it has been noted in Ref. \cite{10Calcagni}, there are several models considered two limited aspects of gravity with fractional derivatives without
embedding them in a fundamental theory: Newtonian gravity and cosmology.
The author of this paper emphasizes that the theories with fractional derivatives do not
have Lorentz symmetry and they are technically difficult due to the presence of fractional
derivatives. We would like to draw the attention of researchers who are interested by scalar theories with fractional derivatives and scalar theories with fractional d’Alembertian to this paper and to "an unconventional review" \cite{2Calcagni}.

In Ref.\cite{1Giusti}, Newton’s potential was derived from an {\it ad hoc} fractional Poisson equation.
Following the hypothesis that the matter distribution of galaxies behaves
as a fractal medium with non-integer dimension, and solving a Poisson equation
with fractional Laplacian,  one can describe the properties of such matter distribution \cite{2Giusti}.

In the paper \cite{Landim},  the fractional dark energy model with  the accelerated expansion
of the Universe driven by a non-relativistic gas  with a
non-canonical kinetic term. It is shown that the inverse momentum operator appears in fractional quantum mechanics and can be expressed through inverse of the Riesz fractional derivative.

In recent paper \cite{Mendoza}, a toy model for extending the Friedmann equations of relativistic cosmology with fractional derivatives is considered.  The cosmological consequences of replacing the integer time derivatives in the Friedmann equations with fractional derivatives (that is using LSM) are explore. This simple approach allowed them to write down the fractional Friedmann equations as follows:
\begin{eqnarray}
\left(\frac{D^{\gamma} a}{D t^{\gamma}}\right)^2=\kappa a^2\left(\frac{ 8\pi
G\rho+\Lambda c^2}{3}\right),\label{18}\\
\frac{D^{\gamma}}{D t^{\gamma}}\left(\frac{D^{\gamma} a}{D t^{\gamma}}\right)=\kappa a\left(-\frac{ 4\pi
G\rho-\Lambda c^2}{3}\right),\label{19}
\end{eqnarray}
for a pressure-less flat Universe. The constant $\kappa$, with dimensions of $time^{2(1-\gamma)}$ has been introduced into
the fractional Friedmann equations in order to have dimensional coherence. The left-hand
side of equation (\ref{19}) is written as such since in general $D^{\gamma}D^{\gamma}\neq D^{2\gamma}$. It is easy to see the obvious difference between the sets of Friedmann equations (\ref{18})-(\ref{19}) and (\ref{1}). Then the authors
introduced Milgrom’s acceleration constant $a_0$ as a fundamental physical quantity for
the description of gravitational phenomena at cosmological scales. With this and since
the velocity of light $c$ and Newton’s gravitational constant $G$ are also fundamental,  it follows from the dimensional analysis that $\kappa = A(a_0/c)^{2(\gamma-1)}$,where $A$ is a dimensionless constant. Due to the  fractional order of the derivative,
one  could adapt the cosmographic parameters as follows. For example, the fractional
Hubble parameter can be defined as:
\begin{equation}
\label{20} H^{*}=\frac{1}{a}\frac{D^{\gamma} a}{D t^{\gamma}}.
\end{equation}
In a matter dominated Universe, where the dark energy density parameter
$\Omega_{\Lambda} = 0$, the following ansatz
is proposed:
\begin{equation}
\label{21} a = a_1t^n,
\end{equation}
where $a_1$ is a constant, and using the rules of fractional
derivative for a power law given in Appendix A, the fractional
Hubble parameter $H^{*}$ is given by:
\begin{equation}
\label{22}
H^{*} =\frac{\Gamma(n + 1)}{\Gamma(n + 1 -\gamma)}t^{-\gamma}.
\end{equation}
As the standard Hubble parameter for the scale factor (\ref{21}) is $H = nt^{-1}$, one can get
\begin{equation}
\label{23}
\frac{H^{*}}{H} =\frac{\Gamma(n + 1)}{\Gamma(n + 1 -\gamma)}t^{1-\gamma}=\frac{\Gamma(n + 1)}{\Gamma(n + 1 -\gamma)}\left(\frac{H}{n}\right)^{\gamma-1}.
\end{equation}
Finally, the equation for the Hubble parameter is obtained as follows
\begin{equation}
\label{24}
H =\frac{a_0}{c}t^{1-\gamma}\Omega_{M}^{1/2(\gamma-1)},
\end{equation}
where $\Omega_M$ stands for the matter density.
Moreover,  in the definition of the fractional cosmographic
parameters $q^{*}, j^{*}$ and $s^{*}$,  the standard Hubble parameter $H$ is used. With the use of
the LSM technique,  $H$ is used for the simplicity.

The authors of Ref. \cite{Mendoza} then apply a fitting procedure to the SN Ia data to estimate the unknown order values of the fractional derivative and fractional cosmographic parameters. It is noted that such a simple approach could explain the current accelerated expansion of the universe without using the dark energy component. Further,  the best fit results for the fractional derivative model with three free parameters are reported. These three parameters are  the
fractional derivative order $\gamma$, the matter density parameter $\Omega_M$ and the power for the scale factor $n$.  They  are presented with their corresponding errors for the initial values  as follows: $n = 2.6,  \gamma = 2.1$ and $\Omega_M = 4.5$. It is interesting that with the mean values mentioned above, the Hubble constant $H_0$ has the following numerical value
$H_0 = 66.95\, km/s\cdot Mpc$, which is in a great agreement with the value reported by Planck \cite{Planck}.

\section{Fractional action cosmology}

\subsection{From FALVA to FAC}

First of all, we have to mention papers \cite {C3, C4}, in which the approach to the dynamical field theories in general and to the theory of gravitation in particular is developed on the basis of the variational principle, formulated by the author, for the action of fractional order (the so-called FALVA). In this approach in the framework of ISA, the action functional integral  $S_L [q] $ for the Lagrangian $L(\tau, q (\tau), \dot q
(\tau)) $ can be written as the fractional integral (A.5):
\begin{equation}
\label {25} S_L [q_i]=\frac {1}{\Gamma
(\alpha)}\int\limits_{t_0}^{t} L(\tau,q_i (\tau),\dot
q_i(\tau))(t-\tau)^{\alpha-1} d\tau .
\end{equation}
At fixed $t$ it is the Stieltjes integral with integrating function $${\displaystyle g_t(\tau)=
\frac{1}{\Gamma(1+\alpha)}[t^{\alpha}-(t-\tau)^{\alpha}]},$$ having
the following scale property:
$
g_{\mu t}(\mu \tau)=\mu^{\alpha}g_t(\tau),~~\mu>0.
$
Then $q_i (\tau) $ satisfies the modified (or fractional)
Euler-Lagrange equation:
\begin{equation}
\label {26}
\frac {\partial L}{\partial q_i}-\frac{d}{d\tau}\Bigl(\frac
{\partial L}{\partial \dot
q_i}\Bigr)=\frac{1-\alpha}{t-\tau}\frac{\partial L}{\partial \dot
q_i}\equiv\overline{F}^i,~~i=1,2,...,n;~~ \tau\in (0,t),
\end{equation}
where the dot over a symbol stands for the first derivative with respect to time $ \tau $, $\overline{F}^i$ is the
modified decaying force of "friction", that is the general
expression for non-conservative force. In the article \cite{D1},
time $ \tau $ is treated as the intrinsic (proper) time, and $t $
is the observer time. The authors of Refs. \cite{C3, C4, C15}
stated that at $ \tau \to\infty $ one has $ \overline {F} ^i=0 $,
and provided some examples of application FALVA to the Riedmann
geometry and perturbed cosmological models.However, these articles do not use FALVA directly for the gravitational action of $ S_G $, written according to (\ref{25}), but try to take into account the influence of the fractional order in action (\ref{25}) on the Friedman equations through the perturbed and time-dependent the classical gravitational constant. Starting with the Lagrangian $L=g _ {ij} (x, \dot x) \dot
x^i\dot x^j $,  the modified geodesic equation has been obtained as:
\begin{equation}
\label {27}
\ddot x^i+\frac{\alpha-1}{T}\dot x^i+\Gamma^i_{jk}\dot
x^j\dot x^k=0,
\end{equation}
where $\Gamma^i_{jk}$ are the usual  Christoffel symbols, and
$T=t-\tau$.  The second term here is interpreted as a dissipative
force, which infinitely increases as $ \tau \to t $ for $ \alpha
\ne 1 $ and under condition of fixing future  time $t$.
Using  non-relativistic approximation,  the time variation of Newton's gravitational constant  and, as a results,   the perturbation of the gravitational constant  have been found as $ \Delta G = {\displaystyle \frac {3
(1-\alpha)} {4\pi \rho T} \frac {\dot a} {a}} $. After that, the so-called effective gravitational constant $G _ {eff} =G +\Delta G $ can be substituted  into the standard Friedmann equations:
\begin{equation}
\frac{1}{a^2}({\dot a}^2+k)= \frac{8\pi G_{eff}}{3}\rho +
\frac{\Lambda}{3}{,}~~~~\frac{\ddot a}{a}= -\frac {4\pi G_{eff}}{3}(\rho+ 3
p)+\frac{\Lambda}{3}{,}\label{28}
\end{equation}
where the cosmological constant $\Lambda$ equals to zero or $\Lambda=
(\beta/t)(\dot a/a)$ \cite{C17}, where $G _
{eff} =G=const $. Then,  the solutions
of equations (\ref{28}) are studied for several barotropic EoS $p =w \rho $ \cite{C17, C18}.

As it noted in Ref. \cite{1Shchigolev}, using the canonical parameter transform $s=g_t (\tau) $ in  equation (\ref{27}), one can reduce it  to the usual geodesic equation:
\begin{equation}
\label {29} \ddot x^i+\Gamma^i_{jk}\dot x^j\dot x^k=0,
\end{equation}
where the over dots stand for derivatives with respects to $s$.
So that means that equations (\ref{28}) could be
obtained without applying FALVA but  by replacing
the parameter $s=g_t (\tau)$.  The same remarks could be address to Ref.\cite{C19},in which the modified  cosmological equations are obtained  using
the periodic weight function $g_t (\tau) $ in the generalized action (\ref{25}) as follows:
\begin{equation}
\label {30} S_L [q_i]=\int\limits_{t_0}^{t} L(\tau,q_i (\tau),\dot
q_i(\tau))\exp(-\chi \sin (\beta \tau)) d\tau .
\end{equation}
The geodesic equation could be obtained  by
variation of (\ref{30}) as
$$
 \ddot x^i-\beta \chi \cos (\beta \tau)\dot
x^i+\Gamma^i_{jk}\dot x^j\dot x^k=0.
$$
The second term was again considered here as
perturbation of the gravitational constant $\Delta G = {\displaystyle
\frac {3\chi \beta \cos (\beta \tau) H} {4 \pi \rho}}$.
Nevertheless, the weight function in this case is defined by $
{\displaystyle \frac {d g_t (\tau)} {d \tau}} = \exp (-\chi \sin
(\beta \tau))$, and the transform  $s=g_t (\tau) $ in (\ref{29}) leads to the same equation.

Similarly to Refs. \cite{C18, C19} , the number of fractional action cosmological models  are considered
in articles \cite{1Debnath}-\cite{Jamil}, where the models are followed again from the fractional action applied to the classical Lagrangian of a curve defined by $L = (1/2)g_{ij}\dot x^i \dot x^j$ , where  $g_{ij}$ is the metric tensor. The modified Friedmann equations are used in the form:
\begin{equation}
H^2+\frac{2(\alpha-1)}{T_1}H+\frac{k}{a^2}= \frac{8\pi G}{3}\rho, ~~~\dot H -\frac{(\alpha-1)}{T_1}H-\frac{k}{a^2}= -4\pi G(\rho+
p),\label{31}
\end{equation}
where $H(t)=\dot a/a$ is the Hubble parameter, and $T_1=t-\tau$. Using equations (\ref{31}),
 a varying gravitational coupling constant, the model of dark energy in this
paradigm and  relevant cosmological parameters are obtained.

It is important to note that, strictly speaking,  the action (\ref{30}) and its generalizations in Refs. \cite{1Debnath}-\cite{Jamil} are  not the fractional ones contrary to  the action functional (\ref{25}) which is really a fractional integral (A.5).
Therefore, using the concept of fractional calculus of variations (or the fractional action-like variational approach, FALVA), the fractional action cosmology should deal  with fractional weight function as it has been proposed  in \cite{C20}-\cite{5Nabulsi}.
Indeed, as it noted in Ref. \cite{1Shchigolev}, it would be more correct to apply FLAVA  directly to the gravitational  fractional action functional $S_G $ trying to  modify the Friedmann equations in this way.
Later on, all models on the basis of such modifications of the cosmological equations were called "Fractional Action Cosmology" or FAC. In FAC,  the action integral $S_L [q]$ in (\ref{25}) with Lagrangian density $L(t', q_i(t'), \dot q_i(t'))$ is represented  as a fractional Riemann-Stieltjes integral: $S^{\alpha}_L [q_i]=\Gamma^{-1}
(\alpha)\int\limits_{t_0}^{t} L(t',q_i (t'),\dot
q_i(t'))(t-t')^{\alpha-1} d t'$ with the integrating function ${\displaystyle g_t(t') = \Gamma^{-1}
(\alpha) [t^{\alpha} - (t-t')^{\alpha}]}$. This approach realizes  the space
scaling concepts of Mandelbrot to define the scaling in fractional time as
$d^{\alpha}t = \pi^{\alpha/2} \Gamma^{-1}(\alpha/2) |t|^{\alpha-1} dt$, where $0<\alpha <1$ \cite{Mandelbrot}.

\subsection{The FAC Models}
\noindent

A lot of  the FAC models could be derived using a variational principle for the modified fractional Einstein-Hilbert action with a varying cosmological term $\Lambda$, that is  $\displaystyle S^{\alpha}_{EH}=M_{P}^2\int \sqrt{-g}\,g_{t'}(t)\,d^{\,4}x (R-2\Lambda)/2$, where $M_P^{-2}=8\pi G$ is reduced Planck mass. First of all, one can suppose that the matter content of the universe is minimally coupled to gravity. Then, the total action of the system is $S_{total}^{\alpha}=S_{EH}^{\alpha}+S_{m}$. Here,  the effective action for matter $S_{m}$can be represented either by $S^{\alpha}_{m}=\int {\cal L}_m\sqrt{-g}\,g_{t'}(t)\, d^4x$, which follows from the fractional matter action similar to (\ref{25}), or by the usual expression for the matter action  with standard measure $S_{m}=\int {\cal L}_m\sqrt{-g} d^4x$.

\subsubsection{FAC Models with a fractional matter action}

The first option is realized in Refs.\cite{1Shchigolev, 2Shchigolev}.  Using definition (\ref{25}), the modified fractional effective action in a spatially flat Friedmann-Robertson-Walker interval (\ref{2}) ,  that is $\displaystyle ds^2 = N(t)^2 dt^2 - a^2(t) \delta_ {ik} dx^i dx^k$, where $N$ is the lapse function and $a(t)$ is a scale factor, can be given by a fractional integral as \cite{1Shchigolev}:
\begin{equation}
\label{32} S^{\alpha}_{eff}=\frac {1}{\Gamma
(\alpha)}\!\int\limits_{0}^{t}\! N \!\left [ \frac{3}{8\pi G}
\left(\frac{a^2\ddot a}{N^2}+\frac{a\dot a^2}{N^2}-\frac{a^2\dot a
\dot N}{N^3}-\frac{\Lambda a^3}{3}\right)+
a^3 {\cal L}_m\!\right
](t-\tau)^{\alpha-1}\! d\tau,
\end{equation}
where $\alpha \in (0,1)$,  and ${\cal L}_m$ is the Lagrangian density of matter represented by the energy density $\rho$ and pressure $p$.
where  such a choice  yields the following  modified continuity equation,
\begin{equation}
\dot \rho+3\left(H+\frac{1-\alpha}{3 t}\right)(\rho+p)=0,\label{33}
\end{equation}
and the set  of the following modified Friedmann equations
\begin{eqnarray}
 3 H^2+3\frac{(1-\alpha)}{t}H=\rho+\Lambda{~,}~~~~~~~~~~~~~~~\label{34}\\
 2\dot H+ 3 H^2+2\frac{(1-\alpha)}{t}H +
\frac{(1-\alpha)(2-\alpha)}{t^2}=-p+\Lambda{~,}\label{35}
\end{eqnarray}
where the gravitational constant $8 \pi G = 1$, and $H(t) = \dot a / a $ is the Hubble parameter.
In the standard FRW cosmology of GR, the continuity equation for a perfect  fluid is  the energy conservation law  for matter which is followed from the Bianchi identity. Therefore,  the continuity equation (\ref{33}) could be derived from the field equations (\ref{34}), (\ref {35}). These equations also yield the modified continuity equation (\ref{32}) in the case $\alpha \neq 1 $ , but only if the following equation is valid \cite{2Shchigolev}:
\begin{equation}\label{36}
\dot H + 3 H^2  - \frac{2(4-\alpha)}{t}H
-\frac{(1-\alpha)(2-\alpha)}{t^2}
= \frac{t \dot \Lambda}{1-\alpha}.
\end{equation}
This final form of the equation is obtained after dividing by $ (1 - \alpha) \neq 0$. Therefore, in the limit of the standard cosmology of general relativity with $ \alpha = 1 $, the equation preceding the equation (\ref {36}) uniquely leads to the constant cosmological term $ \Lambda $, and the set of equations (\ref{34}), (\ref{35}) becomes the usual Friedmann equations.

Several exact models have been obtained on the basis of main FAC equations (\ref{34})-(\ref{36})
in Refs. \cite{1Shchigolev, 2Shchigolev} under the assumption of a vacuum-like state of matter that fills the universe, or on the basis of a rather  general {\it ansatz} for the dynamical cosmological term. In any case, the behavior of the models demonstrate a significant difference from the corresponding standard models. Obviously, this is followed from the fractional nature of the action functional.

From equations (\ref{34}) and (\ref{35}), the effective EoS $w_{eff} = p_{eff} / \rho_{eff} $, where $p_{eff} = p-\Lambda$ and $\rho_{eff} = \rho + \Lambda $, has been proposed  as
\begin{equation}\label{37}
w_{eff}= -1-\frac{2}{3}\,\frac{\displaystyle \frac{\dot H}{H^2}-\frac{1-\alpha}{2(tH)}+\frac{(1-\alpha)(2-\alpha)}{2(tH)^2}}{\displaystyle 1+\frac{1-\alpha}{(tH)}},
\end{equation}
and coincides with the standard expression  $w_{eff} = \displaystyle -1 - \frac{2}{3} \, \frac{\dot H}{H^2}$ in the limit $\alpha \to 1$

The simplest  example of exact solution to equations (\ref{34})-(\ref{36}) has been  found for the quasi-vacuum EoS of matter: $w=-1$.  From equation (\ref{33}), it follows that  $\rho (t) = \rho_0 = constant$ and $p = - \rho = - \rho_0$ similarly to the standard GR cosmology. Then, equations
(\ref{34})-(\ref{36}) can be easily solved for the Hubble parameter,
\begin{equation}
\label{38} H = \frac{C_{\alpha}}{t}+ H_0\,
t^{\displaystyle\frac{1-\alpha}{2}}~,
\end{equation}
where
$C_{\alpha}=\displaystyle\frac{(1-\alpha)(2-\alpha)}{(3-\alpha)}\,$,
$H_0$ is a positive constant of integration, and the scale factor of the universe:
\begin{equation}
\label{39} a = a_0\,\, t^{\displaystyle C_{\alpha}}\exp
\left(\frac{3-\alpha}{2}H_0 t^{\displaystyle \frac{3-\alpha}{2}}
\right) ~.
\end{equation}

Moreover,  a class of exact models is obtained using the solutions of equation (\ref{36}) followed to the assumption that cosmological term $\Lambda(t)$  is a known function of time.  With the help of  the following substitution
\begin{equation}\label{40}
x=\ln (t/t_0) \Leftrightarrow t=t_0 \exp(x);\,\,\, Y(t) = t\, H(t),
\end{equation}
where $t_0> 0$ is a constant, equation  (\ref{36}) can be rewritten as follows
\begin{equation}\label{41}
Y'(x) - (9-2\alpha) Y(x) + 3 Y^2(x) -(1-\alpha)(2-\alpha)=\frac{t_0^2}{1-\alpha}e^{\displaystyle 2x}\Lambda '(x),
\end{equation}
where the prime denotes the derivative with respect to $x$. Taking into account the structure  of this equation, it could be assumed that there exists a class of solutions with the cosmological term satisfied the following equation:
\begin{equation}\label{42}
\Lambda '(x) = \frac{1-\alpha}{t_0^2} e^{\displaystyle -2x}\Big(k_1 Y'(x) + k_2 Y(x)+k_3 Y^2(x) +k_4\Big),
\end{equation}
where $ k_i $ are arbitrary constants.  Finally,  some  examples of exact solutions with the phenomenological functions $\Lambda(t)$,  widely discussed in the literature (see, e.g. Refs.\cite{Overduin, Sahni2}) are obtained using  {\it ansatz}
\begin{equation}\label{43}
\dot \Lambda = (1-\alpha)\Big[ k_1\frac{\dot H}{t}+(k_1+k_2)\frac{H}{t^2}+k_3\frac{H^2}{t}+\frac{k_4}{t^3}\Big].
\end{equation}
Several exact solutions for the FAC models were proposed in Ref.\cite{2Shchigolev} for different values of $k_i$.

\subsubsection{FAC Models with a standard matter action}

The case of the standard matter action  is studied in Refs. \cite{Shchigolev12}-\cite{Shchigolev13}. Using the fractional variational procedure in a spatially flat FRW metric (\ref{2}) for the total action of $\displaystyle S^{\alpha}_{EH}=M_{P}^2\int \sqrt{-g}\,g_{t'}(t)\,d^{\,4}x (R-2\Lambda)/2$ and  $S^{\alpha}_{m}=\int {\cal L}_m\sqrt{-g}\,g_{t'}(t)\, d^4x$, the following dynamical equations can be obtained
\begin{eqnarray}
3 H^2+3\frac{(1-\alpha)}{t}H=t^{1-\alpha} \rho+\Lambda ,~~~~~~~~~~~~~~~~~\\ \label{44}
2\dot H + 3 H^2+\frac{2(1-\alpha)}{t}H +
\frac{(1-\alpha)(2-\alpha)}{t^2}\!=\!-t^{1-\alpha} p+\Lambda ,\label{45}
\end{eqnarray}
where again $8\pi G\, \Gamma (\alpha)=1$ for the sake of simplicity. However, the continuity equation can written in its usual form as
\begin{equation}
\dot \rho+3H(\rho+p)=0,\label{46}
\end{equation}
expressing the standard energy conservation law for a perfect fluid.
Therefore, it can be mentioned that the perturbed continuity equation is not a specific property of the FAC, while it  almost always arises in many modifications of the theory of gravity. The main idea of Ref.\cite{Shchigolev12} is to keep the usual form of the continuity equation within the FAC. As one can see, this aim was achieved using  the concept of fractional order for the action functional only in relation to the gravitational sector. In addition, it was also proposed to write the system of basic equations in such a way that the effective $ \Lambda $ - term could be considered as a kinematically induced (by the Hubble parameter) cosmological term. It was shown on a specific example that a model based on this proposal can lead to some fairly realistic modes of expansion of the Universe.

Using Eqs. (\ref{44}) and (\ref{45}), one can obtain the EoS of matter as follows:
\begin{equation}\label{47}
w_m = \frac{p}{\rho}= -1-\frac{2}{3}\,\frac{\displaystyle \frac{\dot H}{H^2}-\frac{1-\alpha}{2(tH)}+\frac{(1-\alpha)(2-\alpha)}{2(tH)^2}}{\displaystyle 1+\frac{1-\alpha}{(tH)}-\frac{\Lambda}{3 H^2}}.
\end{equation}

Moreover, it can be shown that these equations yield  the continuity equation (\ref{46}) in the case $\alpha \neq 1$, only if
\begin{equation}
\frac{d }{d t}\Big(t^{\alpha - 1}\Lambda\Big)=\frac{3(1-\alpha)}{t^{ 2-\alpha}}\,\Big[\dot H - 2\frac{(2-\alpha)}{t}H\Big]. \label{48}
\end{equation}
This equation can be solved in quadratures as
\begin{equation}
\Lambda(t)=\Lambda_0 t^{1-\alpha}+3(1-\alpha)\left[\frac{H(t)}{t}-(2-\alpha)t^{1-\alpha}\int t^{\alpha-3} H(t)d t\right], \label{49}
\end{equation}
where $\Lambda_0$  is a constant of integration. Substituting Eq. (\ref{49}) into the model equations  (\ref{44}), (\ref{45}), the following set of equations:
\begin{eqnarray}
3H^2=t^{1-\alpha}\rho_{eff},~~~~~~~~~~~~~~~~~~~~~~~~\\ \label{50}
2\dot H+3H^2-\frac{1-\alpha}{t}H+\frac{(1-\alpha)(2-\alpha)}{t^2}=-t^{1-\alpha}p_{eff},\label{51}
\end{eqnarray}
were obtained. Here the effective energy density and pressure are represented by
\begin{equation}
\rho_{eff}=\rho+\Lambda_{eff},\,\,\,\,\,p_{eff}=p-\Lambda_{eff}, \label{52}
\end{equation}
where $\Lambda_{eff}=\Lambda_0-3(1-\alpha)(2-\alpha)\int t^{ \alpha-3}H(t)d\,t$.
The last equation supposes  that the effective cosmological term consist of the cosmological constant $\Lambda_0$ and the induced  cosmological term.

As shown in Ref.\cite{Shchigolev12},  a broad class of exact solutions to the equation (\ref{48}) in the case $w_m \ne -1$ can be obtained using  the following {\it ansatz} for the cosmological term $\Lambda(t)$:
\begin{equation}\label{53}
\Big[ e^{\displaystyle-(1-\alpha)x}\Lambda(x)\Big]' = \frac{3(1-\alpha)}{t_0^2} e^{\displaystyle(\alpha-3)x}\Big[c_1 Y'(x) + c_2 Y(x)+c_3+F(x)\Big],
\end{equation}
where $x=\ln (t/t_0), Y(t) = t\, H(t)$,  $c_i$ and $t_0> 0$ are constants, and $F(x)$ is an arbitrary smooth function. Applying  (\ref{53}) to Eq. (48),  the following expression for the Hubble parameter was obtained:
\begin{equation}\label{54}
H(t) = \frac{1}{t} \cdot \left[\frac{H_0 L}{K} \cdot t^{\displaystyle L/K}  -\frac{M}{L} +\frac{t^{\displaystyle L/K}}{K}\int F(t)t^{\displaystyle -L/K-1}d\,t\right],
\end{equation}
where $H_0$ is an integration constant. The generic character of the generating function $F(t)$ provides a lot of  opportunities in constructing  the exact models. So, in the simplest  case $F(t)\equiv0$, the scale factor of the model follows hybrid law of evolution:
\begin{equation}\label{55}
a(t) = a_0 t^{\displaystyle -M/L} \exp \left\{H_0 t^{\displaystyle L/K}\right\}
\end{equation}
where $K = 1-c_1,\,L = 5-2\alpha +c_2,\,M = c_3$.
Using the same approach, a number of exact solutions were obtained in Ref. \cite{Shchigolev11}.

\subsection{Testing FAC}

As  assumed in Ref.\cite{Shchigolev13} in order to determine the evolution of FAC model,  the following effective cosmological term widely  discussed in the literature (see, e.g., Refs.  \cite{Overduin, Sahni2}) could be used
\begin{equation}\label{56}
\Lambda_{eff} = \beta H^2.
\end{equation}
where $\beta =\frac{3}{2} m H_0^{1-\alpha}$, $m, H_0$ are constants, and $\alpha \in (0,1)$.
In terms of the dimensionless cosmic time $\tau = H_0 t$, the following solution to the given FAC was obtained
\begin{equation}\label{57}
H(\tau)=H_0\left[\frac{1-\alpha}{m \tau^{2-\alpha}}+\frac{m-(1-\alpha)}{m}\right],
\end{equation}
Thus, in the far future, the Hubble parameter (\ref{56}) tends the following value
$$
H_{\infty}\equiv H(t \to \infty)=\frac{ (m-1+\alpha)}{m}H_0,
$$
so that $\Lambda_0=\beta H_{\infty}^2$. From equation (\ref{56}), the scale factor of this model can be found as
\begin{equation}\label{58}
a(\tau) \!=\! a_0 \exp \left[\Big(1\!-\!\frac{1-\alpha}{m}\Big)\tau\!-\!\frac{1}{m \tau^{ 1-\alpha}}\!+\!\frac{2\!-\!\alpha-m}{m}\right],
\end{equation}
where $a_0 = a(\tau=1)$. Because the red shift $z$ is defined as $1+z=a_0/a(\tau)$, one can obtain it  from equation (\ref{58}) as
\begin{equation}\label{59}
z\!=\! \exp \left[\frac{1}{m \tau^{1-\alpha}}-\Big(1-\frac{1-\alpha}{m}\Big)\tau-\frac{2-\alpha-m}{m}\right]-1.
\end{equation}

The deceleration parameter $q = -a^2\, \ddot a/\dot a^2 = -1-\dot H/H^2$ in FAC is defined just as in the standard cosmology, since it is a cosmography parameter of the model, and  can be represented as
\begin{equation}\label{60}
q(\tau)=-1+\frac{m(1-\alpha)(2-\alpha)\tau^{1-\alpha}}{[1-\alpha+(m-1+\alpha)\tau^{2-\alpha}]^2}.
\end{equation}
In the paper \cite{Shchigolev13}, this model was subjected primarily to theoretical diagnostics based on cosmographic parameters and the so-called $Om$ diagnostics \cite{C8}. This made it possible to analyze the behavior of the model for various values of its main parameters. An even more important result of the cited article lies in the numerical estimates of the model parameters obtained from observational data.

Since this model contains three independent parameters ( $\alpha$, $m$ and $H_0$),  the observational constraints on all these parameters can be done using 28 data points of $H(z)$  in the redshift range $0.07 \le z \le 2.3$ \cite{Farooq, Busca}.
The observational data consist of measurements
of the Hubble parameter $H_{obs}(z_i)$  at redshifts $z_i$ , with
the corresponding one standard deviation uncertainties $\sigma_{Hi}$.
 is well consistent with the observation result from Planck+WP \cite{Planck} :
$H_0 = 67.3 ± 1.2 \,\, km s^{-1}Mpc^{-1}$.

Then, using the best fit values of the main cosmographic parameters for the Union 2.1 SNIa data from Table II in Ref. \cite{Capozziello2},
\begin{equation}
H_0=69.97^{+0.42}_{-0.41}\,\, km s^{-1}Mpc^{-1},~~~
q_0=-0.5422^{+0.0718}_{-0.026},~~~r_0=0.5762^{+0.4478}_{-0.3528}.\label{61}
\end{equation}
the following values can be obtained: $\alpha \approx 0.926,\,\,~~~m \approx 0.174$,
with the same accuracy as the cosmographic parameters  given by (\ref{61}).

\subsection{Some recent studies of FAC}

In  Refs. \cite{Nabulsi, 4Nabulsi},   the FRW cosmology characterized by a scale factor
obeying different independent types of fractional differential equations was studied, and both types of fractional operators: the Riemann-Liouville fractional integral and the Caputo
fractional derivative were considered.   The solutions for such models are given in terms of Mittag-Leffler and generalized Kilbas-Saigo-Mittag-Leffler functions.

In Ref.\cite{Nabulsi112} the wormholes solutions based on FAC are studied, and  the cosmic dynamics in the presence of wormhole  in closed FRW universe are discussed. As it found, the cosmic acceleration with traversable wormhole may be realized without the need of exotic matter  unless the scale factor of the universe obeys a power law dominated by a {\it negative} fractional parameter which is constrained from SNe Ia data.

Fractional action cosmology with variable order parameter was constructed in
Ref. \cite{Nabulsi113}, where a large number of cosmological equations are
obtained depending on the mathematical type of the fractional order parameter. This idea  results on a number of cosmological
scenarios  and their dynamical consequences. It was observed that the used fractional cosmological formalism is able to create a large family of solutions and offers new features not found in the standard formalism and in many fundamental research papers.

The late-time evolution of a flat FRW model of the universe in the context of a non-minimal fractional cosmology characterized by fractional weight in time was studied in paper \cite{6Nabulsi} . It was shown that due to  the conformal coupling between the scalar field and gravity, a negative time-dependent quadratic potential and provided that the scale factor of the universe is related to the scalar field through a power-law ansatz, the universe is oscillating with time. Moreover, the oscillating behavior of the EoS parameter can be realized around  -1 by crossing the phantom divide line an infinite number of times.

The short communication \cite{Nabulsi114}proposed an original approach  based on a generalized fractional integral operator which mixes the Riemann-Liouville
and the Erdelyi-Kober integrals in one single operator
usually “the Glaeske-Kilbas-Saigo fractional”.
This generalized fractional integral is defined by
$${}_a I^{\alpha,\beta,\gamma}f(x) = \frac{1}{\Gamma(\alpha)}\int\limits_{a}^{x} f(t)(x-t)^{\alpha-1}e^{-\beta t}e^{-\gamma (x-t)} dt,
$$
where $(\alpha,\beta,\gamma)$ are constants. A number of non-singular gravitational fields  are obtained without using extra-dimensions, and some examples is provided to show that these gravitational fields hold many motivating features in space-time
physics.

A significantly different approach to the application of fractional derivatives in cosmology from the above approach is presented in Ref. \cite{Torres} , where quantization is studied in terms of the fractional derivative of the cosmological coupling theory of the non-minimal derivative, namely, the  Fab Four John theory. Its Hamiltonian version is the problem of fractional powers of momenta. Moreover, this problem is solved using the so-called conformable fractional derivative \cite{Khalil}, which leads to the Wheeler-DeWitt equation of the second order. It has been shown that a wide range of scale factor solutions are possible, including a bouncing solution.

The prospect of using such local fractional derivatives as Conformable Fractional Derivatives (CFD) in cosmology and astrophysics is confirmed by work \cite{Yousif}, where the fractional equation of an isothermal gas sphere is written and solved. to conformable fractional isothermal gas spheres

In this study,  the fractional form of isothermal Lane-
Emden for the CFD isothermal
gas sphere is considered using  the power series method and obtain a recurrence
relation for the power series coefficient. In order to evaluate the fractional parameter
impact on the configuration of the stars, the physical parameters
of the isothermal gas sphere are derived and determined for the neutron
stars.

One more new approach to FAC is presented in  the recent article \cite{Rasouli} , where  a brief summary of fractional quantum mechanics is given  in order to motivate towards fractional quantum cosmology.  A model of stiff matter in a spatially flat homogeneous and isotropic universe is investigated and  discussed. A new quantum cosmological solution, where fractional calculus implications
are explicit, is presented and then contrasted with the corresponding standard quantum cosmology setting.

\section{Conclusion}

Summing up, we can conclude that in this short review we have presented the main results of studies on cosmological models with fractional derivatives aimed at attempts to explain the accelerated expansion of the universe outside the hypothesis of the existence of exotic types of matter. Indeed, there are not too many such attempts and the results of these attempts are  not so significant, which is primarily due to the difficulties in justifying the admissibility of the use of fractional calculus in the cosmology  and gravity. This is indeed a problem, despite some success and advances of fractional analysis in  several fields of science such as engineering, biology and so on.  Despite this, based on the achievements of the application of fractional analysis to the problems of gravity, certain really successful results in fractional cosmology have been obtained, presented in this review. It was noted that such studies have certain achievements and perspectives, possibly related to the new definition of fractional derivatives, such conformal fractional derivatives and some others. We hope that this review will be useful in discussing this problem and the results obtained will find their worthy application to subsequent cosmological studies.

\appendix
\section{Fractional  derivatives and integrals}

Today there are more than two dozen definitions of the fractional derivative \cite {C1}. In physical and technical applications of fractional differential calculus, the Riemann-Liouville derivative (RLD), the Caputo derivative (CD) and some others are most applicable.

Such derivatives are defined by means of analytical continuation
of the Cauchy formula for the multiple integral of integer
order as a single integral with a power-law core into the field of
real order $\mu>0$:
\begin{equation}
\label{appeqn}
{}_c I^{\mu}_x f(x)=\frac{1}{\Gamma(\mu)}
\int\limits_{c}^{x} f(t)(x-t)^{\mu-1} dt.
\end{equation}
The Riemann-Liouville derivative of fractional order $\alpha\ge
0 $ of function $f(x)$ is defined as the integer order derivative
of the fractional-order integral (\ref {1}):
\begin{equation}
\label{appeqn}
D^{\alpha}_x f(x)\equiv D^n_x {}_c I^{n-\alpha}_x
f(x)=\frac{1}{\Gamma(n-\alpha)}\frac
{d^n}{dx^n}\int\limits_{c}^{x} \frac{f(t)}{(x-t)^{\alpha-n+1}}dt
\end{equation}
where $D^n_x \equiv d^n/d x^n$,~~$n=[\alpha]+1$. This definition
corresponds to the so-called left derivative, frequently denoted as
${}_c D^{\alpha}_x f(x)$. For the limit $\alpha=1$, this
definition gives $df(x)/dx$. For example, the left RLD of $x^k$
for $\alpha\le 1, c=0$ equals:
\begin{equation}
\label{appeqn}
D^\alpha_x
x^k=\frac{\Gamma(k+1)}{\Gamma(k+1-\alpha)}x^{k-\alpha}.
\end{equation}
For $\alpha=1$, one has the usual result: $D^1_x x^k=k x^{k-1}$.
The interesting feature of RLD is that RLD of non-zero constant
$C_0$ does not equal zero, but for $\alpha\le 1$ it equals
$D^\alpha_x C_0=C_0 x^{-\alpha}/\Gamma(1-\alpha)$. The right RLD
is defined similarly to (\ref{2}) on the interval $[c,d]$:
\begin{equation}
\label{appeqn}
 {}_x D^{\alpha}_d f(x) =
\frac{1}{\Gamma(n-\alpha)}\left(-\frac {d}{dx}\right)^n
\int\limits_{x}^{d} \frac{f(t)}{(t-x)^{\alpha-n+1}}dt
\end{equation}
It should be emphasized again that the Riemann-Liouville fractional integral of order $\alpha$ is defined by
\begin{equation}
\label{appeqn}
I^{\alpha} f(x)=\frac{1}{\Gamma(\alpha)}
\int\limits_{c}^{x}(x-t)^{\alpha-1} f(t) dt,
\end{equation}
and has a memory kernel.

One needs to be aware that according to the formulas of addition of
orders, the following holds(see \cite{C1}, p.161):
$$
D^{\alpha}_xD^{\beta}_x f(x)= D^{\alpha+\beta}_x f(x)-\sum^n_{j=1}
D^{\beta-j}_x
f(c+)\frac{(x-c)^{-\alpha-j}}{\Gamma{(1-\alpha-j)}}~,
$$
that is $D^{\alpha}_xD^{\beta}_x f(x)\ne D^{\alpha+\beta}_x f(x)$,
if only not all derivatives $D^{\beta-j}_x f(c+)$ at the beginning
of the interval are equal to zero. That is why
$D^{\alpha}_xD^{\alpha}_x f(x)\ne D^{2\alpha}_x f(x)$ in the general
case. Generalizing the Laplace operator in the equation for Newtonian
gravitational potential, the author of \cite{C5} wrongly doubles
the order of the repeated fractional derivative. The authors of
\cite{C11} have avoided this mistake, having written down the
Laplacian $\Delta^{\alpha}$ as:
$$
\Delta^{\alpha}u=\frac{1}{r^{2\alpha}}D^{\alpha}_r(r^{2\alpha}D^{\alpha}_r
u)+\frac{\Gamma^2(\alpha +1)}{r^{2\alpha} \sin^{\alpha}\theta}
\frac{\partial}{\partial\theta}(\sin^{\alpha}\theta \frac{\partial
u}{\partial\theta})+\frac{\Gamma^2(\alpha +1)}{r^{2\alpha} \sin^{2
\alpha}\theta} \frac{\partial^2 u}{\partial \phi}.
$$
One can note one more property of the fractional derivative
expressed in modification of the Leibniz rule (see \cite{C1},p.162):
\begin{equation}
\label{appeqn}
D_x^{\alpha}\left[f(x)g(x)\right]=\sum^{\infty}_{k=0}\frac{\Gamma(\alpha+1)}{k!\Gamma{(\alpha-k+1)}}
D_x^{\alpha-k}f(x)D_x^k g(x),
\end{equation}
which becomes the usual rule as $\alpha=n$. It can be represented
as the integral over the order of fractional derivative:
$$
D_x^{\alpha}\left[f(x)g(x)\right]=\int\limits^{\infty}_{-\infty}\frac{\Gamma(\alpha+1)}
{\Gamma(\mu+1)\Gamma(\alpha+1-\mu)}D_x^{\alpha-\mu}f(x)D_x^{\mu}g(x)d{\mu}.
$$
These rules of fractional differentiation can lead to an
essential modification to the cosmological models with fractional
derivatives.

However, there are other definitions of fractional derivatives that differ significantly from those given above.
For example, the definitions of fractional derivatives as fractional powers of derivative operators are provided in \cite{Tarasov}.  For this end, the Taylor series and Fourier series are used to define fractional power of self adjoint derivative operator.

Recently, the authors of \cite{Khalil} have defined a new well-behaved simple fractional derivative called "conformable" fractional derivative depending just on the basic limit definition of the derivative.

\end{document}